\documentclass{optica-article}

\journal{opticajournal} % for journals or Optica Open

\articletype{Research Article}

\usepackage{lineno}
\begin{document}
\pagenumbering{arabic}

\title{Electro-Optic Comb Generation Via Cascaded Harmonic Modulation}

\author{Todd Eliason,\authormark{1} Payton A. Parker,\authormark{1,2} and Melanie A. R. Reber\authormark{1,2,*}}

\address{\authormark{1}Department of Chemistry, University of Georgia, Athens, GA 30602}
\address{\authormark{2}Department of Physics and Astronomy, University of Georgia, Athens, GA 30602}
\email{\authormark{*}mreber@uga.edu} %% email address is required; see note below about the corresponding author designation

% use {asbstract*} to suppress the copyright line. Copyright information will be added in production

\begin{abstract*} 
Electro-optical modulation of a continuous wave laser is a highly stable way to generate frequency combs, gaining popularity in telecommunication and spectroscopic applications. These combs are generated by modulating non-linear electro-optic crystals with radio frequencies, creating equally spaced side-bands centered around the single-frequency seed laser. Electro-optic frequency comb architectures often choose between optical bandwidth (cascaded GHz combs) or higher mode density (chirped RF generation). This work demonstrates an electro-optic frequency comb with > 120 GHz of bandwidth and a 75 MHz repetition rate. The comb has three cascaded electro-optic modulators driven at sequentially lower harmonics, the last megahertz modulation dictating the repetition rate. This architecture can modulate at any individual harmonic and repetition rate without changes to the components. This comb can be used in any applications where a stable and tunable repetition rate is needed.

\end{abstract*}

%%%%%%%%%%%%%%%%%%%%%%%%%%  body  %%%%%%%%%%%%%%%%%%%%%%%%%%
\section{Introduction}
Frequency combs were first developed for frequency metrology, as they can measure optical frequencies in the RF domain using RF electronics\cite{Diddams_PRL2000,Diddams_TIM2001}. Aptly named, comb lasers consist of equally-spaced, narrow-linewidth, laser lines forming the comb structure in the frequency domain\cite{Ye_combbook2005}. Frequency comb lasers have found use in precision measurement from metrology to high-resolution spectroscopy, including absorption spectroscopy, photoacoustic spectroscopy, and others \cite{picque_natphotonics_2019, coddington_optica_2016, qiao_optlett_2021, debecker_optexpress_2005, truong_natphotonics_2013, zhuang_laserphotonicsrev_2023}. 

The most common way to generate a frequency comb laser is by using a pulsed, mode-locked laser and stabilizing the carrier-envelope phase. Mode-locked lasers require significant ultrafast laser expertise and a controlled laboratory setting to operate. An alternative to mode-locked lasers is to generate a frequency comb using a continuous-wave (CW) laser modulated with an electro-optic modulator (EOM), known as EOM combs. EOM combs take full advantage RF electronics to generate and stabilize the frequency comb. EOM Combs offer advantages in arbitrary waveform generation, ranging, optical communications, astronomical spectrum correction, and spectrum detection, due to their highly stability and reproducibility compared to mode-locked lasers \cite{zhuang_laserphotonicsrev_2023}. EOM combs can be generated by placing EOMs either inside the laser cavity or external to the laser cavity. The first kind of EOM comb were intracavity EOM Combs. One of the first examples is from Bell Laboratories in 1963, which used a single modulator in a Fabry-Perot cavity \cite{gordon_belltechjour_1963}.  Cavity based EOM combs typically have GHz repetition rates \cite{imai_jquantelec_1999, xiao_optexpress_2008, yang_applopt_2020, kim_biomedoptexpress_2020}. Some intracavity EOM combs utilize micro-resonators on chips to either create or amplify the modulation \cite{he_natcommun_2023, lambert_communphys_2023, wu_applopt_2021}.

Alternatively, the EOM can be external to the laser cavity, used in this work, and has the advantage of separating out the laser gain dynamics from the comb generation. Extracavity EOM combs consist of one or more EOMs in series or in parallel after the CW laser output. Some EOM combs consist of one EOM modulated with a single sine wave at GHz frequencies\cite{barreiro_scirep_2023, zhang_communphys_2023, murata_jseltopquantumelelectronics_2000, morohashi_optlett_2008}. These single EOM combs have limited bandwidth due to number of modes than can be generated from a single EOM without exceeding the RF damage threshold. A GHz repetition rate is often used to maximize the optical bandwidth of these EOM combs, and consequently most are made with waveguide-type EOMs. Other examples of EOM combs use multiple EOMs in series, a cascaded architecture, modulated at a single frequency. Modulators in series generates broader comb because successive modulations act on the comb teeth generated previously, resulting in an overall larger bandwidth. Cascaded EOM combs often have intensity modulators along with phase modulators, driven at the same RF frequency (>10 GHz typically)\cite{zhang_optexpress_2023, kikkawa_electronlett_2023, song_scirep_2022, ishizawa_scirep_2016, carlson_science_2018, kowligy_optlett_2020}.

EOM combs are also generated by driving the EOM with complex waveforms. Chirped RF generation is a method where a swept sine wave is sent to an single EOM. While the repetition rate can be as low as kilohertz, these combs are limited to a few gigahertz in bandwidth\cite{long_optlett_2023, long_physreva_2016}. The current limitations of these architectures lie in the highest frequency of the function generator and the maximum RF power of the EOM. Chirped sine waves can also be used in series and in a cavity to generate combs\cite{kim_biomedoptexpress_2020}. Another method of modulating is called pseudo-random modulation, where a modified square wave drives the EOM creating a comb, albeit with similar limitations of chirped RF generation\cite{wilson_physrevappl_2018, hebert_physrevappl_2016, hebert_optexpress_2015}.

In this paper we demonstrate a comb design that will combine the large bandwidth of single frequency RF modulation with smaller comb tooth spacing to create a comb without the limitations of the single frequency modulation combs and the complex waveform modulation combs. By diving EOMs in series with sequentially lower harmonic frequencies, we take advantage of the bandwidth of the higher repetition rate, while filling in the comb with the smaller repetition rate. There are a few examples in the literature of EOM-based combs using multiple EOMs driven at different frequencies. An early example involved an EOM in a Fabry-Perot cavity, which was driven with a higher harmonic of the Fabry-Perot modes \cite{saitoh_photontechlett_1996}. There have also been EOM combs which contain two micro-resonators in series, first modulated at a higher frequency then modulated at a lower frequency harmonic \cite{buscaino_jlightwavetech_2020, gong_optica_2022}. These examples are all cavity-based modulations and not purely EOM-based. There are other configurations of multiple EOMs driven at different frequencies used to make a tailored optical frequency output, some with the higher frequency EOM first\cite{li_science_2014, li_optexpress_2021, xu_optlett_2021} and some with the lower frequency EOM first\cite{vanhowe_optlett_2004, kashiwagi_optexpress_2016}, which provides necessary background for combining multiple EOMs. 

Since the EOM comb in this work is driven by multiple frequencies, phase-locked loops (PLL) are used to frequency and phase-lock the frequency sources to a stable reference frequency. Deacon et al. utilized PLLs at 25 GHz and 26 GHz to drive separate cascaded EOM Combs, consisting of an intensity modulator EOM in series with a phase modulator EOM\cite{deakin_optlett_2021}. The found the phase stability of a PLL to decrease the noise of dual comb spectroscopy, providing further motivation for utilizing PLL architecture for driving the combs. To the best of our knowledge, this is the only other laser using PLLs to stabilize the frequency and phases of the driving frequencies. 

The comb in the current work has a > 120 GHz comb bandwidth and 80 MHz comb tooth spacing (repetition rate), which creates a comb that will be useful in spectroscopy and metrology. Like many other EOM-based combs, this one could also be further broadened in a nonlinear fiber and amplified. The organization of the paper is as follows. Section \ref{archtecture} describes the EOM comb in detail, section \ref{combdesign}, and the methods used to characterize the comb, section \ref{characterization}. The tests of the RF electronics are found in section \ref{RFcharacterizationsection} and the optical comb characterization is in section \ref{lasercharacterizationSection}. A discussion of the sideband intensities is found in section \ref{intensitysection} and the paper is summarized in section \ref{summary}.

\section{Architecture}
\label{archtecture}

\begin{figure}[t!]
\centering\includegraphics[width=12cm]{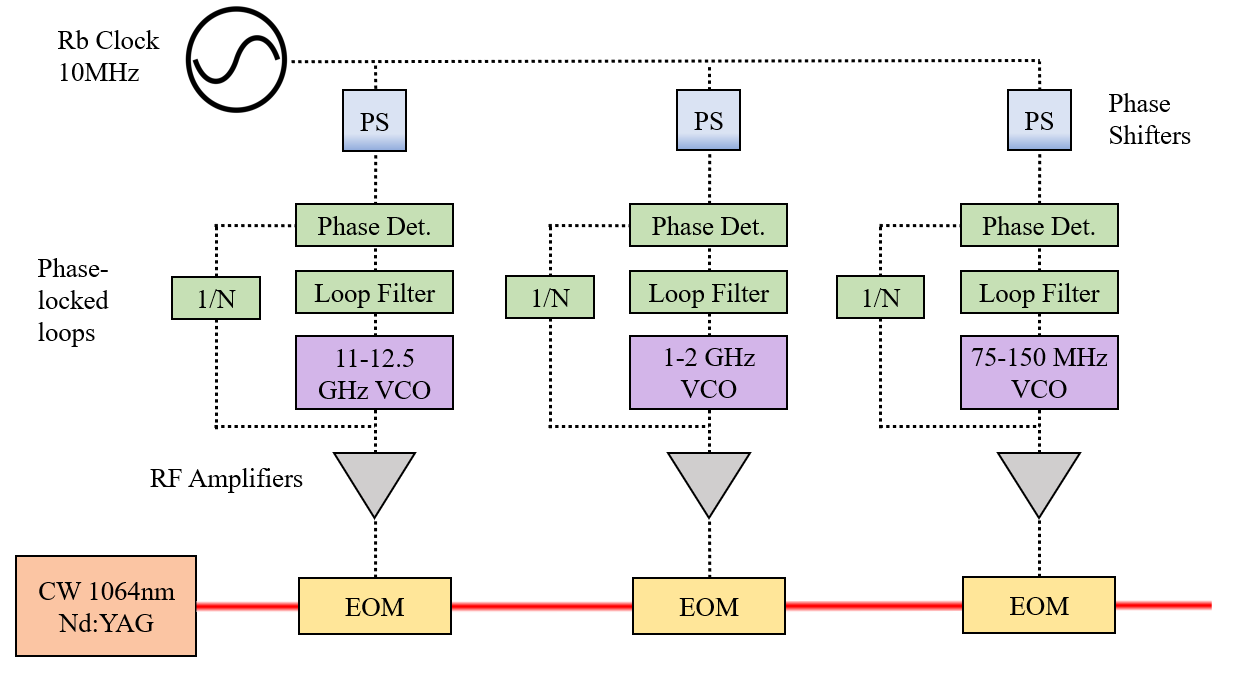}
\caption{EOM Comb Architecture and electronic block diagram, full details in text. PS is a phase shifter, Phase Det. is the phase detector or phase comparator, VCO is a voltage controlled oscillator, and 1/N is a frequency divider.}
\label{comblayout}
\end{figure}

\subsection{EOM Comb Design}
\label{combdesign}

The EOM comb uses a cascaded architecture consisting of three phase modulators in series, pictured in Fig. \ref{comblayout}. The seed laser is a CW Nd:YAG (Coherent Mephisto) with $\le 3$ kHz optical linewidth at 1064 nm. The seed light is coupled into polarization-maintaining fiber and sent through three EOMs in series, starting with the one driven at the highest frequency, and subsequent EOMs driven at a lower frequency. The EOMs are all fiber-coupled, waveguide-type with polarization-maintaining couplers that join the fibers from each EOM in series. The diagnostics and characterization optics are all done in free space. 

The frequency sources for the EOMs are phase and frequency stabilized to a single, stable frequency source through the use of phase locked loops (PLLs). The reference frequency is a 10 MHz Rubidium (Rb) clock (SRS SIM940). Three phase shifters (Synergy Microwave PK-721S) are set before the PLLs to provide relative phase control of each driving frequency. The PLLs are comprised of a phase frequency detector or phase detector, a loop filter, voltage controlled oscillator (VCO), and a frequency divider. The VCO is set to the desired frequency then output to both an amplifier and a frequency divider. The frequency division achieves frequency matching between the VCO frequency that drives the EOM and the Rb clock and is adjustable. The 75-150 MHz PLL  (Analog Devices ADF4152HV) includes as the phase comparator and loop filter and interfaces with an external VCO (Minicircuits ZOS-150+). The 1-2 GHz PLL uses the same PLL board but the VCO is the on-board VCO. The 11-12.5 GHz PLL uses a different PLL (Analog Devices ADF41020), again utilizing the onboard VCO. The PLL output voltages are amplified individually using: 11-12.5 GHz (MiniCircuits ZVE-3W-183+), 1-2 GHz (Minicircuits ZVA-183WX-S+), and 75-150 MHz (Minicircuits ZHL-2010+).  The amplified frequencies are then sent to their respective EOM. The three EOMs used are: 11-12.5 GHz (EOspace PM-DS5-10-PFA-PFA-106-LV), 1-2 GHz (IXBlue NIR-MPX-LN-10 0-10 GHz), and 75-150 MHz (IXBlue NIR-MPX-LN-2 0-2 GHz). The frequencies are set to be exact harmonics of the lowest frequency, within the precision of the frequency divider control. The frequency spacing of the EOMs was chosen to maximize the sidebands produced by each EOM while staying below the RF damage threshold of the EOM, with 13 harmonics between the two lowest frequencies and 11 harmonics between the highest two frequencies.

\begin{figure}[t!]
\centering\includegraphics[width=13cm]{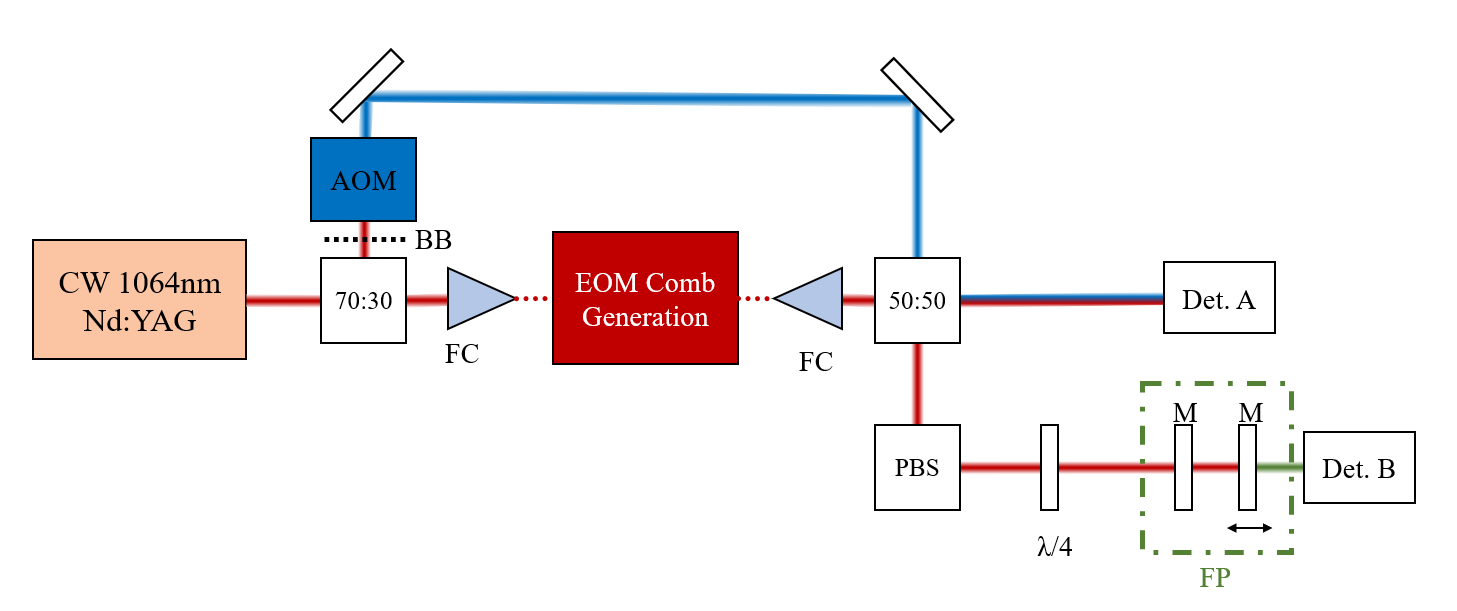}
\caption{Detection Scheme. 70:30 and 50:50 are beam splitters of the same ratio. PBS is a polarizing beam splitter. BB is a beam block that is placed during Fabry-Perot spectra collection. BD is a beam dump. FC are fiber couplers. FP is the Fabry-Perot Cavity. Det A. collects the Heterodyne measurement data. Det. B. collects the Fabry-Perot Spectra. }
\label{diagnostics}
\end{figure}

\subsection{Comb Characterization Methods}
\label{characterization}

To characterize the comb spectra in the RF, both in frequency and phase, it is necessary to beat the comb against another frequency source to separate each comb tooth in frequency to individually detect it on an RF spectrum analyzer. Comb frequencies are rigorously described the carrier envelope offset frequency, $f_0$ and the repetition rate, $f_{rep}$ such that the optical frequency of each $n^{th}$ comb tooth is $v_n = f_0+nf_{rep}$. Then comb is beat with a frequency shifted portion of the CW Nd:YAG laser, $v_{YAG}+f_{AOM}$. The comb teeth are unidentifiable with their distinct beat frequencies of the $n_{th}$ comb tooth, $b_n$ by $b_n = v_n-(v_{YAG}+f_{AOM})=f_0+nf_{rep}-(v_{YAG}+f_{AOM})$. The 80 MHz modulation and the first mode of the 1.04 GHz modulation fall within the bandwidth of the RF spectrum analyzer, which is 1.5 GHz. The 11.44 GHz data was obtained by dividing the signal by eight.

The optical layout is shown in Fig. \ref{diagnostics}). To detect each comb tooth, a portion of the CW Nd:YAG laser is split off and sent to an AOM driven at 260 MHz (Isomet M1250-T260L-0p45 and 536F-L). The frequency shifted Nd:YAG is then made co-linear with the EOM comb and focused onto a detector (Coherent ET-3000) for heterodyne detection. The resultant RF spectrum is amplified with an low noise amplifier (Pasternack PE15A1007), and collected by a RF spectrum analyzer (Rigol DSA815) for frequency measurements and a second RF spectrum analyzer for phase noise analysis (Advantest R3267).

To quantify higher frequency modulations and the EOM comb optical spectra, a homebuilt scanning Fabry-Perot spectrometer is used, as shown in Fig. \ref{diagnostics}. The Fabry-Perot consists of two highly reflective planar mirrors (R > 0.9995, Edmund Optics 89-452) mounted with one mirror on a translation stage controlled by a stack piezo (Thorlabs PC4GR). The piezo is controlled by a high voltage amplifier (Thorlabs MDT694B) and a function generator (Rigol DG1022Z). The light was detected with a large mode area detector after the second mirror (Det B: Thorlabs PDA10A2). The spectra was collected and saved with an oscilloscope (Sigilent SDS2352X-E). A quarter waveplate and polarizing beamsplitter placed before the Fabry-Perot act as an optical isolator preventing the reflected light from heading back to the laser. 

The Fabry-Perot was designed to observe the high frequency modulations and collect the full optical spectrum of the EOM comb. The finesse (>500 typ.) was verified by measuring the line-width of the Nd:YAG, which has a specified line-width $\le 3$ kHz, much lower than the FWHM of the cavity. In our work this means at the largest FSR to capture the full spectral bandwidth, we were not able to resolve individual comb teeth. The free spectral range was varied from as low as 2 GHz, to measure the 80 MHz spectra, to about 175 GHz, for the full spectrum.

\section{Results and Discussion}

\subsection{RF Electronics Characterization}
\label{RFcharacterizationsection}

\begin{figure}[t!]
\centering\includegraphics[width=13cm]{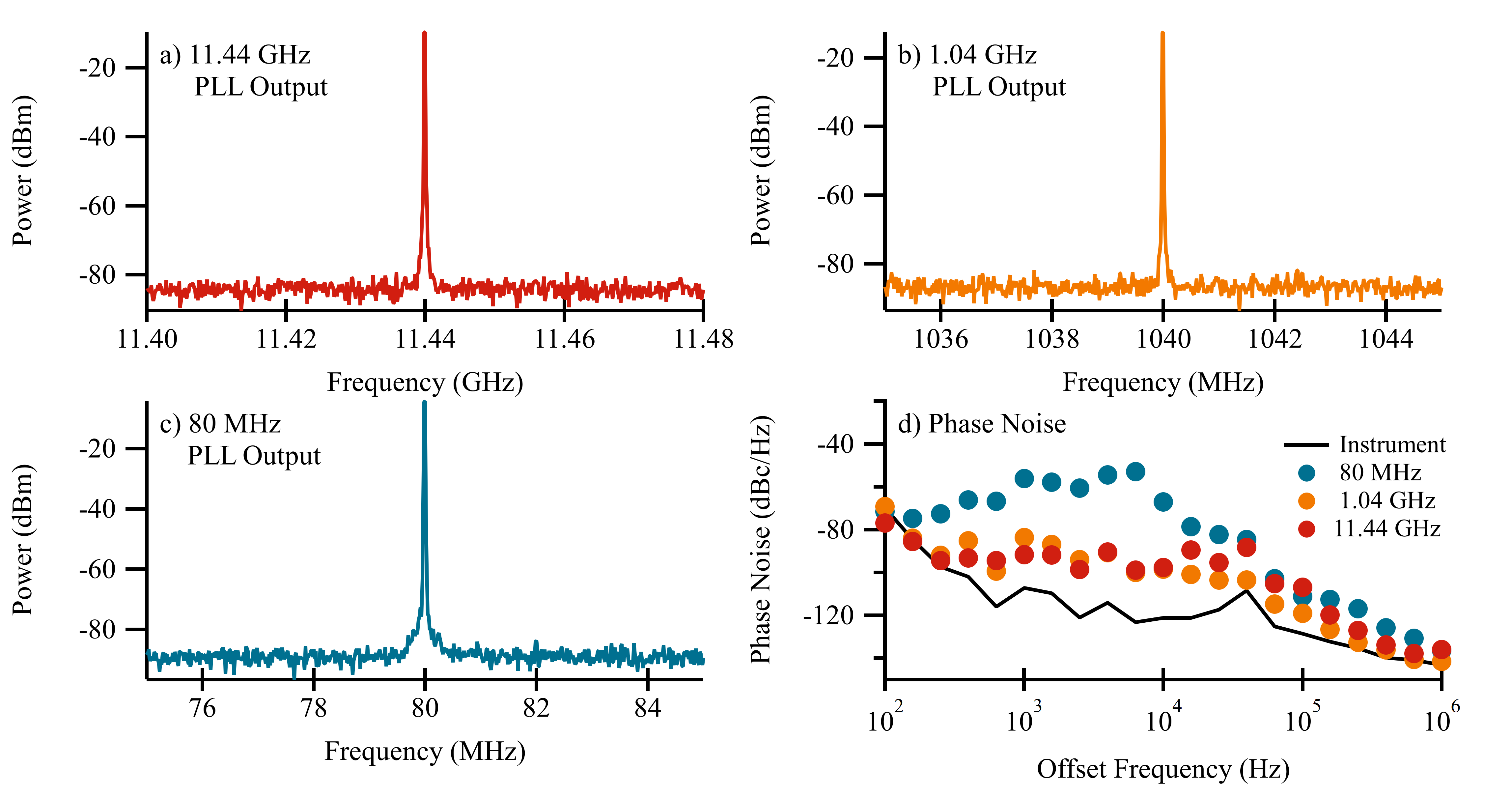}
\caption{a) RF Spectra of the 11-12.5 GHz PLL at 11.44 GHz b) RF Spectra of the 1-2 GHz PLL at 1.04 GHz c) RF Spectra of the 75-150 MHz PLL at 80 MHz d) Phase noise of the RF sources. The 11.44 GHz source was divided by 8 to collect this data.}
\label{RFsources}
\end{figure}

Figure \ref{RFsources} shows the RF spectra, panels a) - c), and phase noise, panel d), at the output of each PLL, with the phase noise collected after amplification. The PLLs are providing the necessary stable, driving frequencies for the EOMs, at the set desired frequencies. There are no additional frequencies being sent to the EOMs, with the exception of some intensity in the higher harmonics. The phase noise of the 80 MHz output is higher than the other PLL outputs, until approximately 100 kHz. This is perhaps not surprising since that PLL is not specifically optimized for 80 MHz, while the PLLs for the other two frequencies were optimized for the driving frequency used. The instrument response of the phase noise was taken with the Rb clock.

\subsection{Laser Characterization}
\label{lasercharacterizationSection}

\begin{figure}[ht!]
\centering\includegraphics[width=13cm]{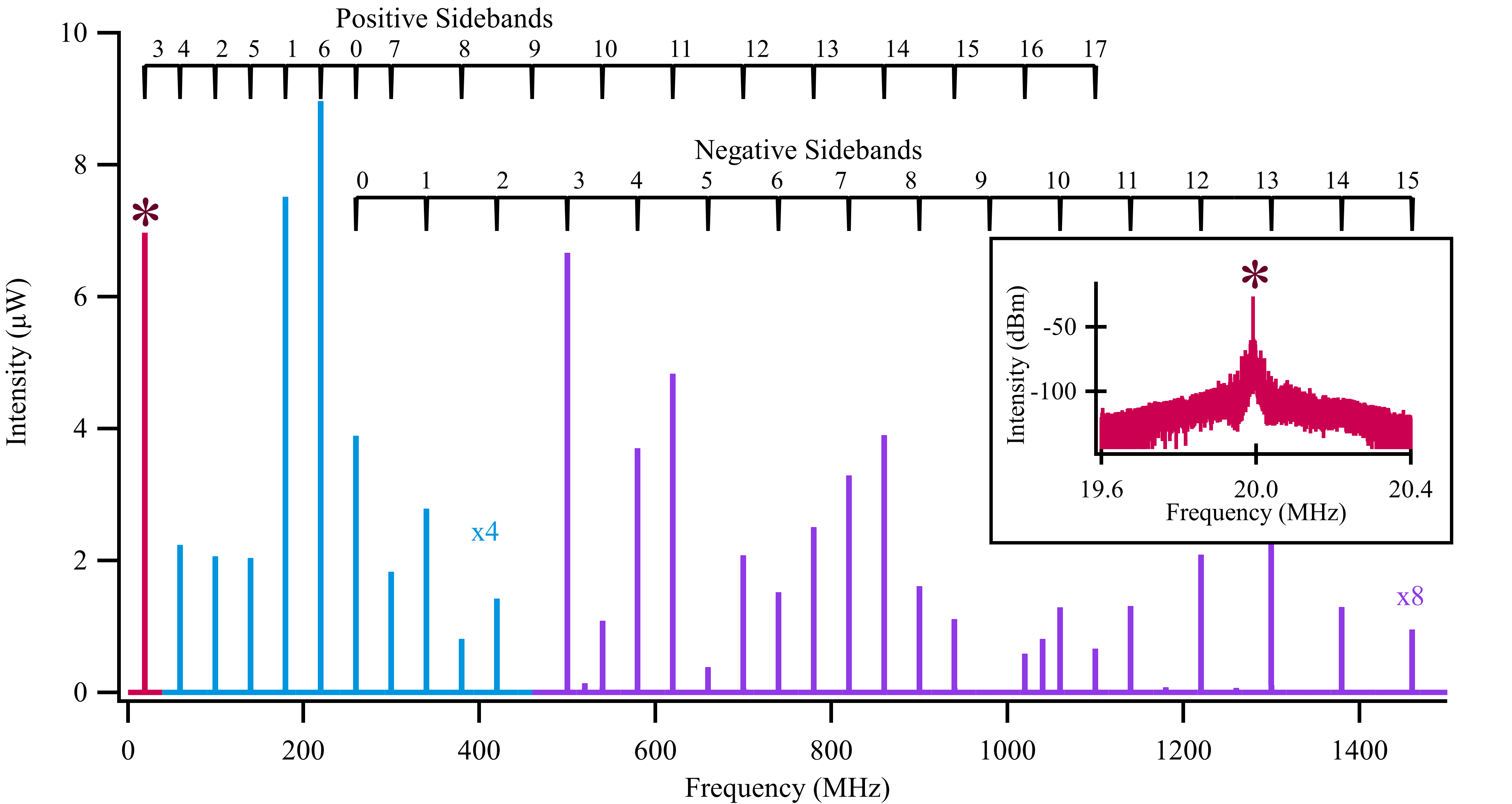}
\caption{RF Spectra of the Nd:YAG seed laser with only the 80 MHz EOM Modulation on beat with the 260 MHz AOM modulated laser. Racetracks denote the positive and negative modulations referenced from the seed laser. The unmodulated seed laser is the $0^{th}$ mode. The light blue section of spectra is multiplied by 4, the purple section multiplied by 8, and the 1/f noise removed for clarity.}
\label{RF80MgOpticalspectrum}
\end{figure}

The RF spectra of the laser is obtained by recording the heterodyne beat with the 260 MHz AOM-shifted seed laser. Figure \ref{RF80MgOpticalspectrum} shows the RF spectra of the seed laser and 80 MHz modulation only. What intensity remains of the unmodulated seed laser, labeled as the $0^{th}$ sideband, shows up at the 260 MHz modulation frequency. The 80 MHz modulation creates 17 observable modes in the positive direction, and at least 15 in the negative direction. This is sufficient modulation to span the 1.04 GHz sidebands, corresponding to the $13^{th}$ sideband from the preceding EOM. The inset zooms in on the $+3^{rd}$ sideband on a log scale so the noise and peak shape are visible. The peaks visible in the spectra and not assigned as peaks from the 80 MHz modulation are harmonics of the AOM driver.

\begin{figure}[t!]
\centering\includegraphics[width=13cm]{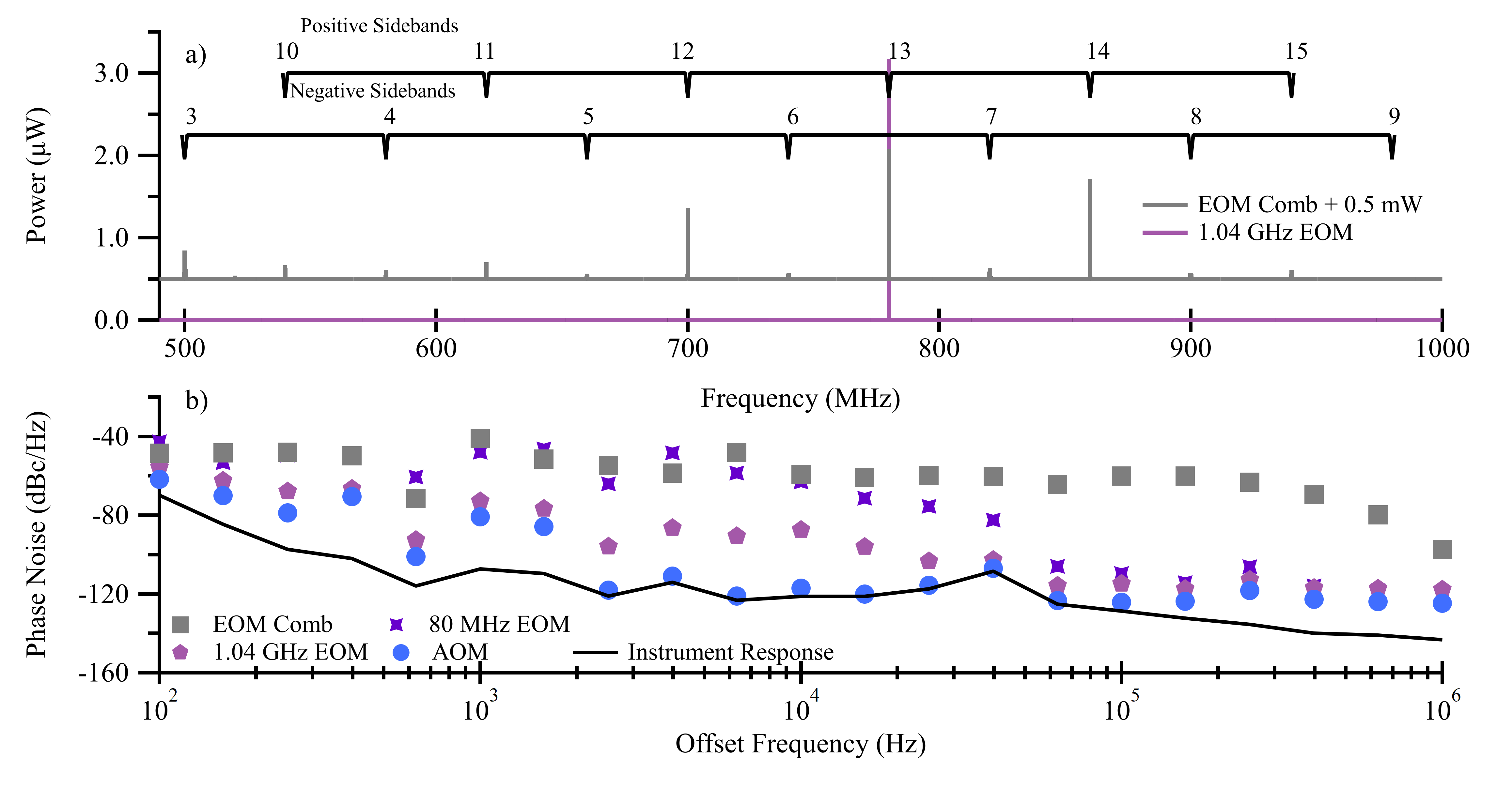}
\caption{RF Spectra a) Selection of the RF Spectrum from 490-1000 MHz, The EOM comb signal is shifted up by 0.5 mW. The racetracks denote the comb tooth number from the seed laser for the positive and negative sides of the EOM comb. b) Phase noise of the instrument (black line), AOM modulated seed laser (blue circle), single modulation at 1.04 GHz (pink circle) and 80 MHz (purple square), and the full comb (gray square)}
\label{phasenoise}
\end{figure}

Figure \ref{phasenoise} a) is the AOM beat RF spectra for the Nd:YAG with just the 1.04 GHz EOM turned on (pink), and for the full comb with all three EOM's turned on (gray). The full EOM comb is, offset by 0.5 mW for clarity and the sideband numbers labeled with the racetracks. The "$0$" label corresponds to the unmodulated Nd:YAG laser frequency, $\nu_{YAG}$, such that the positive $3^{rd}$ comb tooth frequency, for example, is at 240 MHz ($\nu_{YAG} + 240 MHz$) and shows up at 20 MHz in the heterodyne RF beat spectrum. The pink trace shows the optical beat spectra in the RF when only 1.04 GHz EOM is turned on. The positive $13^{th}$ sideband overlaps with the 1.04 GHz modulation, as expected. Note that with the linear scale the noise is not visible.

The phase noise is reported in Fig \ref{phasenoise}  b), of the full EOM comb, the AOM modulated Nd:YAG laser, 1.04 GHz modulation only, 80 MHz modulation only, and the instrument response. The phase noise of the full comb is greater from 1 kHz to 600 kHz that in either the 80 MHz or 1.04 GHz sources. Since we could not directly measure the phase noise of the 11.44 GHz modulation with our spectrum analyzer, the increase in phase noise can possibly be attributed to the 11.44 GHz and/or optical phase noise. The RF bandwidth of the spectrum analyzer (1.5 GHz), detector (2 GHz), and LNA (3 GHz) limit the amount of comb teeth that are observable. To measure the higher frequency modulations, the home-built Fabry-Perot spectrometer was used.

\begin{figure}[t!]
\centering\includegraphics[width=13cm]{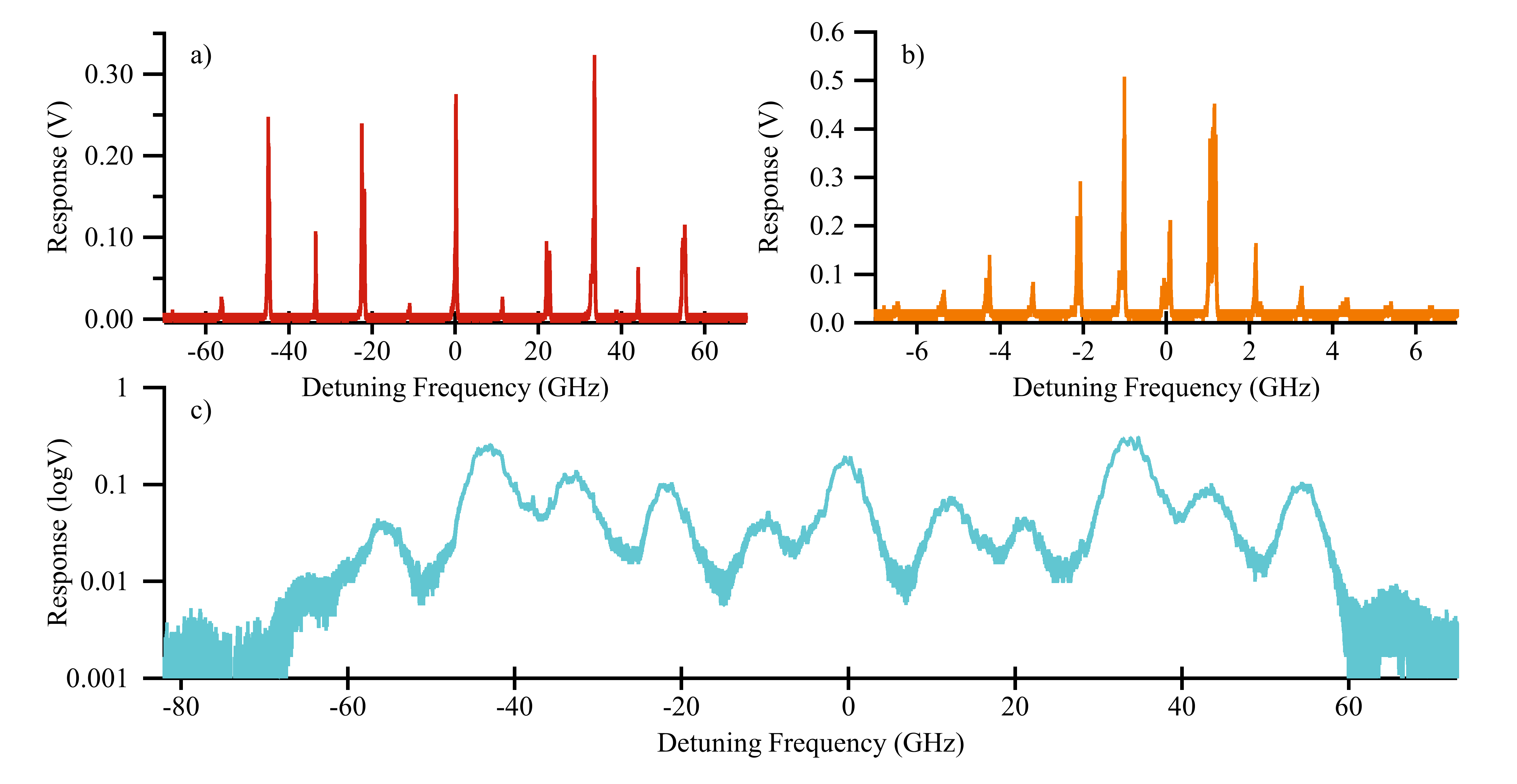}
\caption{Fabry-Perot Spectra a) seed laser modulated at 11.44 GHz  b) seed laser modulated by 1.04 GHz c) The spectrum collected of the full EOM comb, averaged 3 times.}
\label{FabryPerotspectrum}
\end{figure}

The full spectra of the EOM comb, 11.44 GHz modulation, and 1.04 GHz modulations were recorded with the scanning Fabry-Perot spectrometer, as seen in figure \ref{FabryPerotspectrum}. Fig. \ref{FabryPerotspectrum} a) shows the modulation with only the 11.44 GHz EOM on. The first modulation at 11.44 GHz creates 5 frequency modes on either side of the fundamental, which gives the EOM comb its optical bandwidth. Fig. \ref{FabryPerotspectrum} b) is the Fabry-Perot spectra of the laser with only the 1.04 GHz EOM. The intensity profile of each modulation is discussed in section \ref{intensitysection} and consistent with EOM modulation of a single frequency and the first few harmonics, as expected. Fig. \ref{FabryPerotspectrum} c) shows the Fabry-Perot spectrum of the entire EOM comb laser. Individual comb teeth are not resolved, as expected, however intensity is shown across >120 GHz of bandwidth. This optical bandwidth is much greater than many EOM combs with Megahertz or lower repetition rates, before external broadening \cite{long_optlett_2023, long_physreva_2016, wilson_physrevappl_2018, hebert_physrevappl_2016, hebert_optexpress_2015}.

\begin{figure}[ht!]
\centering\includegraphics[width=13cm]{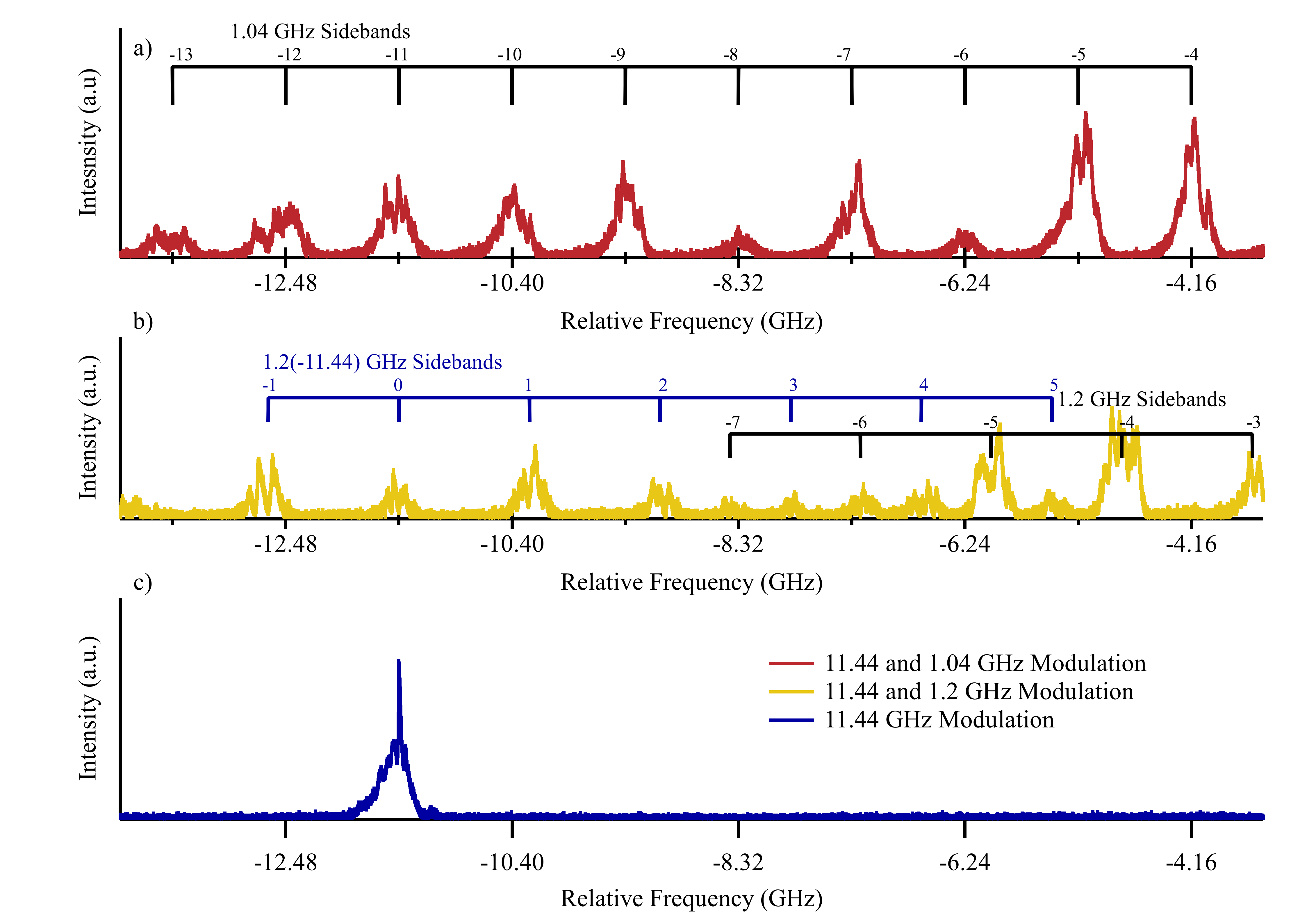}
\caption{Fabry-Perot Spectra a) Trace of 11.44 and 1.04 GHz modulation. The harmonic numbers are measured from the seed laser which is set to 0 GHz. b) Trace of the 11.44 and 1.2 GHz modulation. The black racetrack is counting 1.2 GHz modes starting at the seed laser. The blue racetrack is counting modes generated from the -11.44 GHz mode. The peak width is limited by the Fabry-Perot cavity and is not a measure of the comb tooth width.}
\label{harmonicoverlap}
\end{figure}

To demonstrate the overlap of the sidebands as a result of the harmonic modulation frequencies, Fabry-Perot spectra are taken with and without harmonic modulation for comparison, Figure \ref{harmonicoverlap}. The EOM comb with only 1.04 GHz and 11.44 GHz modulation, the 11$^{th}$ harmonic, are shown in Figure \ref{harmonicoverlap}a) with racetracks counting sidebands from the unmodulated peak, set as 0 GHz. The $-11^{th}$ 1.04 GHz sideband overlaps with the first 11.44 GHz sideband, as expected. The line-shape of each frequency mode is from the scanned Fabry-Perot and not the line-shape of the EOM comb. The laser with only 11.44 GHz EOM modulation is shown in panel~c) for reference. EOM modulation with 1.2 GHz and 11.44 GHz, not a harmonic frequency, is shown in panel~b). The 1.2 GHz modulation sidebands on adjacent 11.44 GHz sidebands do not overlap or fill in the gap between 11.44 GHz modulations. The intensities of the harmonically overlapped peaks compared to the non-overlapping peaks between the 11.44 GHz modulations shows the second advantage of the harmonic modulation scheme. The intensities of the sidebands from adjacent 11.44 GHz modulations adds in the harmonic modulations scheme, especially clear when comparing the "-7" sideband from panel~a) to the "-7 " sideband in panel~b).

\begin{figure}[t!]
\centering\includegraphics[width=13cm]{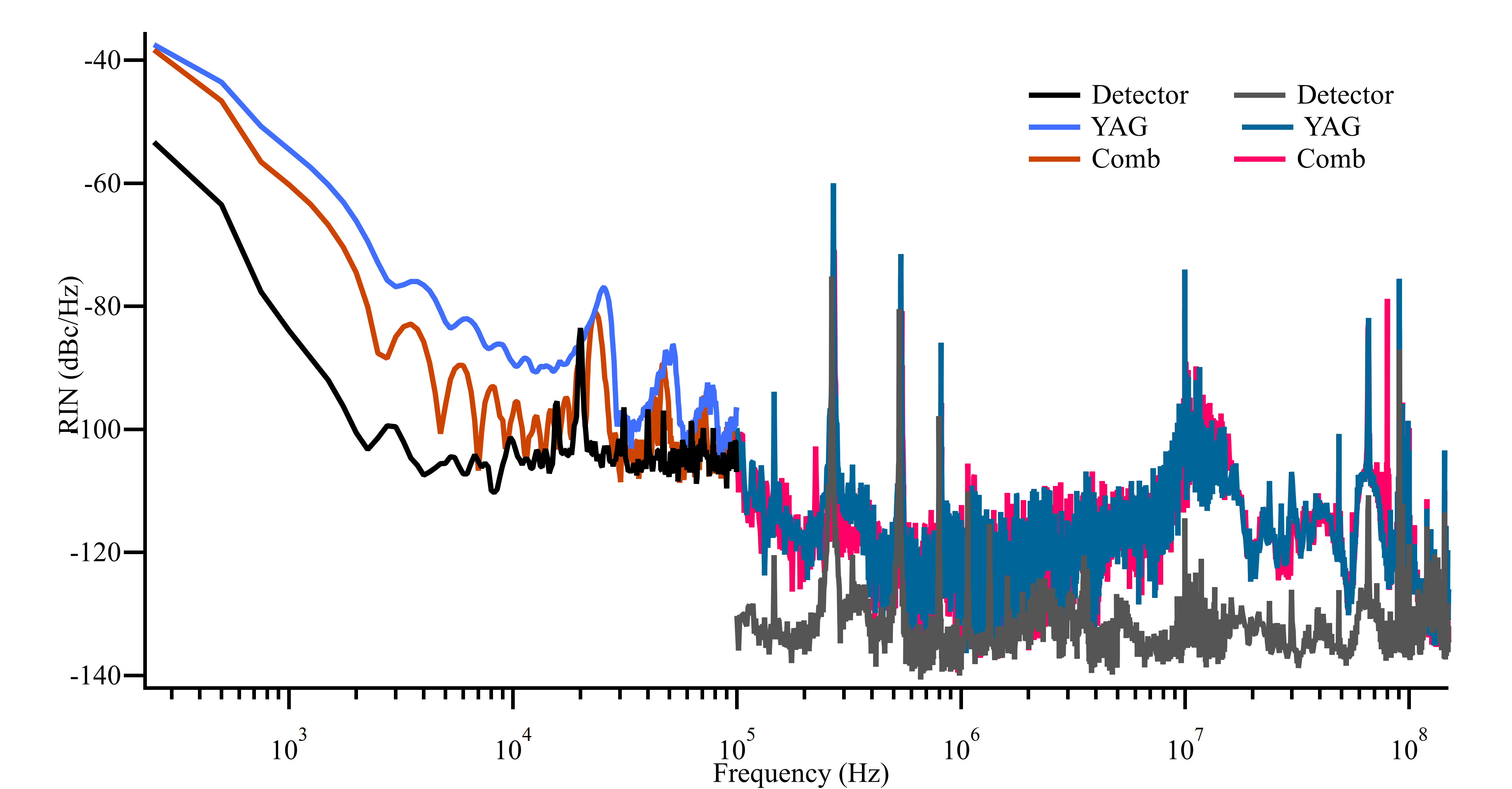}
\caption{RIN collected of the background, YAG seed laser, and the EOM Comb. For frequencies higher than 1 kHz, a low-noise amplifier (LNA) was used after the detector to increase the signal above the detector noise floor. The RIN is dominated by the Nd:YAG seed laser.}
\label{RINfigure}
\end{figure}

The EOM modulation does not seem to appreciably affect the relative intensity noise (RIN) (see Figure \ref{RINfigure}). There are various frequency spurs in the seed laser RIN, and are not a result of the EOM comb generation. The baseline is not continuous because the data is stitched using different amplifiers, which changed the noise floor; the higher frequency data starting at 1 kilohertz used a LNA and a different spectrum analyzer. Since the RIN is unaffected by the EOM comb generation, lower noise seed laser can be used to reduce the RIN.

\subsection{Sideband Intensity}
\label{intensitysection}

The conventional discussion of EOM comb generations uses the Bessel function to describe the intensity of the side-bands generated by a single phase modulation at frequency $\Omega$. This modulation is symmetric about the seed laser:
\begin{equation}
    \begin{aligned}
    E_{out} & = Acos(\omega t-\delta sin(\Omega t))\\
            & =A [ J_0(\delta)cos\omega t + J_1(\delta)cos(\omega+\Omega)t \\
            & + J_1(\delta)cos(\omega-\Omega)t+ J_2(\delta)cos(\omega+2\Omega)t \\
            & + J_2(\delta)cos(\omega-2\Omega)t+ J_3(\delta)cos(\omega+3\Omega)t \\
            & + J_3(\delta)cos(\omega-3\Omega)t+ J_4(\delta)cos(\omega+4\Omega)t \\
            & + J_4(\delta)cos(\omega-4\Omega)t+ ...]
    \end{aligned}
\end{equation}
$E_{out}$ is the electric field of the modulated seed laser\cite{yariv_1985}. The amplitude of each mode is described by Bessel function, $J_n(\delta)$, the $n$ being the side-band number and $\delta$ is the phase modulation index. The phase modulation index is defined as $\delta = (1/2)\pi (V/V_{\pi})$. This solution does not account for multiple frequencies in a single phase modulator. The driving modulations in this work included higher order harmonics inherent to the frequency sources. The output electric field in this work is described in equation 2. The equation can be extended to have any amount of harmonics.
\begin{equation}
\begin{aligned}
    E_{out} = Acos[ \omega t & - \delta_1 sin(\Omega t+\phi_1) \\
    & - \delta_2 sin(2\Omega t+\phi_2) \\
    & - \delta_3 sin(3\Omega t+\phi_3) \\
    & - \delta_4 sin(4\Omega t+\phi_4) ...]
\end{aligned}
\end{equation}
The phase modulation index changes with frequency because $V_{\pi}$ is frequency dependant. This model of the electric field predicts asymmetric side-bands strengths, with the intensities depending upon the phase modulation index of each harmonic and the phase of each harmonic. This predicted intensity pattern is consistent with the asymmetric sideband observed in figures \ref{FabryPerotspectrum} and \ref{RF80MgOpticalspectrum}.

\section{Summary}
\label{summary}

A fully adjustable EOM comb has been demonstrated in this paper. The novel RF generation containing three separate PLLs allows for driving cascaded EOMs at rigorous harmonics of each other creating a broad EOM comb with a MHz repetition rate. The advantage of this architecture is that the repetition rate and harmonics chosen are tunable without any change to the hard components. The EOM comb in this paper can have a final repetition rate from 75 MHz to 150 MHz. The comb generation is all fiber such that this method can be applied to any optical regime where you have a seed laser with acceptable line-width and noise. The line intensity analysis in this work is limited by the noise and linewidth on our homebuilt, scanning Fabry-Perot interferometer used to measure the sidebands. A quantitative analysis of the comb tooth intensities would be a useful extension of this work. This EOM comb is uniquely suited for use in spectroscopy due to the long term stability of the frequency sources combined with the stability of a commercial Nd:YAG and the 80 MHz repetition rate. 

\begin{backmatter}
\bmsection{Funding} This material is based upon work supported by the National Science Foundation under Grant No. 2207784.

\bmsection{Acknowledgments} T.E. would also like to thank U.S. Department of Energy, Office of Science, Office of Basic Energy Sciences, Gas Phase Chemical Physics Program Award Number DE-SC0020268 for funding.

\bmsection{Disclosures} The authors declare no conflicts of interest.

\bmsection{Data availability} Data underlying the results presented in this paper are not publicly available at this time but may be obtained from the authors upon reasonable request.

\end{backmatter}

\bibliography{EOMcomb.bib}

\end{document}